\documentclass[fleqn,twoside]{article}
\usepackage{espcrc2}
\usepackage{graphicx}

\newcommand{\AmS}{{\protect\the\textfont2
  A\kern-.1667em\lower.5ex\hbox{M}\kern-.125emS}}
\title{QGSJET-II: results for extensive air showers}

\author{S.~Ostapchenko\address[MCSD]{Institut f\"ur Experimentelle Kernphysik,
University of Karlsruhe, 76021 Karlsruhe,  Germany}%
        \address{D.V. Skobeltsyn Institute of Nuclear Physics, 
         Moscow State University, 119992 Moscow, Russia}%
        \thanks{Now at Forschungszentrum Karlsruhe, Institut f\"ur Kernphysik,
76021 Karlsruhe,  Germany.}%
\thanks{ This work has been supported in part  by the German Ministry
for Education and Research (BMBF, Grant 05 CU1VK1/9).}}

\begin{document}

\begin{abstract}
The new hadronic Monte Carlo model QGSJET-II is applied for  
extensive air shower (EAS) calculations. The obtained results are  
compared to the predictions of the original QGSJET and of the SIBYLL 2.1 
interaction models. It is shown that non-linear  effects
change substantially model predictions for hadron-nucleus interactions
and produce observable effects for calculated EAS characteristics.
Finally the impact of the  
new model on the interpretation of air shower array data is 
discussed.
\vspace{1pc}
\end{abstract}

\maketitle

\section{INTRODUCTION}

Experimental studies of high energy cosmic rays (CR) are mainly performed
on the basis of extensive air shower (EAS) techniques:
measuring various characteristics of nuclear-electro-magnetic cascades,
 induced by primary CR particles in the atmosphere, one uses the obtained
 knowledge to reconstruct the properties of the initial particles. 
 Because of a very complicated structure
 of the atmospheric cascade such an analysis necessarily requires
 both particle detectors -- to measure various EAS components at Earth
 or in space, and proper simulation tools, which allow to mimic the cascade
 process and to establish a connection between the measured information
 and the primary particle characteristics. An important part of EAS
 simulation programs are Monte Carlo (MC) models of hadronic interactions.
 In the absence of a rigorous theoretical approach for a treatment of
 general (minimum-bias) hadron-nucleus and nucleus-nucleus collisions
 a development of reliable hadronic MC generators for very high energy
 interactions is far non-trivial; also due to the necessity to extrapolate
 corresponding experimental knowledge from the energies of present colliders
 to much higher CR energies. On the other hand, measuring different EAS
 components allows to perform a cross-check of model description and to
 discriminate between available MC generators.
 
 During last years QGSJET model \cite{kal93,kal94,kal97} has been 
 successfully used in the field, in particular, being employed in popular 
 EAS simulation programs,  CORSIKA \cite{hec98} and AIRES \cite{sci99}.
 Also model tests, performed for example by the KASCADE collaboration,
 using its multi-component EAS detector setup \cite{ant03}, showed that
 the model reproduces measured shower characteristics and their correlations
 reasonably well \cite{ant99,mil04}. However, the mentioned agreement is
 still far from being ideal; recent investigations showed some discrepancies
 in the description of basic correlations between electron and muon EAS
 components \cite{ulr04}. Also annoying is the fact that analyzing different
 EAS components results in sizably divergent conclusions concerning primary
 CR composition, for all models in use \cite{hor02,wat04}.
 
 Recently a new hadronic MC model QGSJET-II has been developed
 \cite{pylos1}, which included a treatment of non-linear 
 interaction effects in individual hadronic (nuclear) collisions.
 The latter gave a possibility to obtain a consistent description of
 interaction cross sections and parton momentum distributions in hadrons,
 compared to corresponding measurements, thus solving the basic deficiency
 of the original QGSJET and providing a much more reliable model
 extrapolation into ultra-high energy domain. Additionally, realistic nuclear
 density parameterizations have been employed in the model, individually
 for each nucleus \cite{kal99}, more reliable two-component low mass
 diffraction treatment has been used  \cite{kal93}, and all the model 
 parameters have been re-calibrated using a wider set of accelerator data.
 The goal of the present work is to investigate the impact of the mentioned
 modifications on the calculated air shower characteristics and to draw 
 possible consequences for EAS data interpretation.

\section{RESULTS FOR HADRON-NUCLEUS INTERACTIONS}

Despite the fact that the physics of hadronic interactions is very
complicated and contributes to air shower development in a rather non-trivial
way, basic EAS observables mainly depend on a limited number of macroscopic
characteristics of hadron-air collisions. Indeed, the most fundamental
EAS parameter, the position of the shower maximum $X_{\max}$ is grossly
defined by  inelastic hadron-air 
cross sections and by the interaction inelasticities, the latter quantity
being defined as the relative difference between the lab. energies of the
 initial and the most energetic final particles. While the number of charged
 leptons $N_e$, measured at a given observation level, is strongly
 correlated with $X_{\max}$, this correlation is much weaker
 for corresponding number of muons $N_{\mu}$, and even becomes negative
 for relatively high muon threshold.
 On the other hand, muon numbers appear to be quite sensitive to 
 multiplicities of elementary interactions.

\begin{figure}[t] 
\begin{center}
  \includegraphics[width=7cm,height=3.8cm,angle=0]{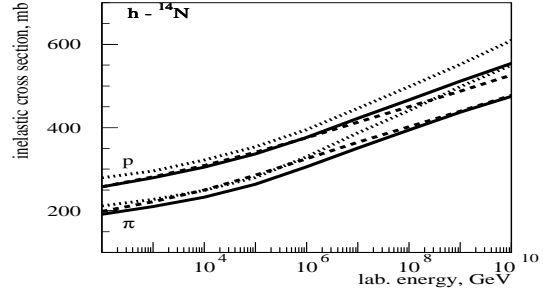} 
\end{center}
\vspace*{-1.3cm}
   \caption{Inelastic hadron-air cross sections for QGSJET-II, QGSJET-I,
   and SIBYLL 2.1 models - smooth, dashed, and dotted
   curves.\label{sigair}} 
\end{figure}

\begin{figure}[t] 
\vspace*{-.4cm}
\begin{center}
  \includegraphics[width=7cm,height=3.8cm,angle=0]{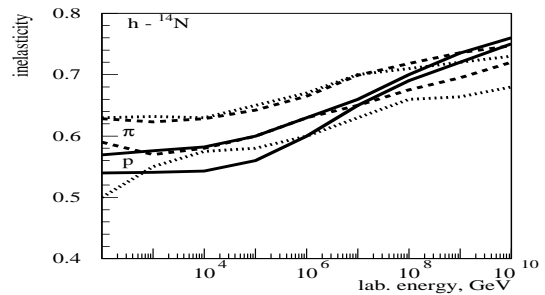} 
\end{center}
\vspace*{-1.3cm}
   \caption{Inelasticities of  hadron-air interactions for QGSJET-II, QGSJET-I,
   and SIBYLL 2.1 models - smooth, dashed, and dotted
   curves.\label{kinair}} 
\end{figure}

\begin{figure}[htb] 
\begin{center}
  \includegraphics[width=7cm,height=4cm,angle=0]{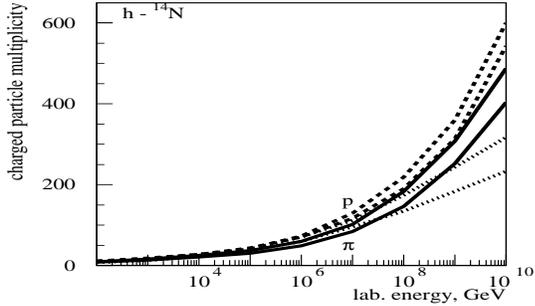} 
\end{center}
\vspace*{-1.3cm}
   \caption{Multiplicities of charged
particles in  hadron-air interactions for QGSJET-II, QGSJET-I,
   and SIBYLL 2.1 models - smooth, dashed, and dotted
   curves.\label{mulair}} 
\end{figure}

 In Figs. \ref{sigair}--\ref{mulair}
  the predictions of the new model for inelastic cross sections
 $\sigma_{h-\rm{air}}^{\rm{inel}}$, 
 inelasticities  $K_{h-\rm{air}}^{\rm{inel}}$, and multiplicities of charged
particles $N_{h-\rm{air}}^{\rm{ch}}$,   are plotted in comparison with
 both  the original QGSJET and the SIBYLL 2.1
  \cite{eng99} models. Compared to the original model, the results of the
   QGSJET-II are defined by a competition of two effects: non-linear
   screening corrections, which lead to a reduction of the interaction
   eikonal and correspondingly of the number of elementary particle
   production processes (cut Pomerons), and steeper parton momentum
   distributions, leading to a faster energy increase of the latter
   quantities \cite{pylos1}. At not too high energies the first effect
   dominates, leading to smaller numbers of $\sigma_{h-\rm{air}}^{\rm{inel}}$, 
  $K_{h-\rm{air}}^{\rm{inel}}$, and $N_{h-\rm{air}}^{\rm{ch}}$,
  especially for pion-air interactions. On the other hand, in the very high
  energy limit the influence of parton distributions prevails and the new
  model predicts larger values for the quantities of interest.


\section{AIR SHOWER CHARACTERISTICS}

\begin{figure}[t] 
\vspace*{-.4cm}
\begin{center}
  \includegraphics[width=7cm,height=3.9cm,angle=0]{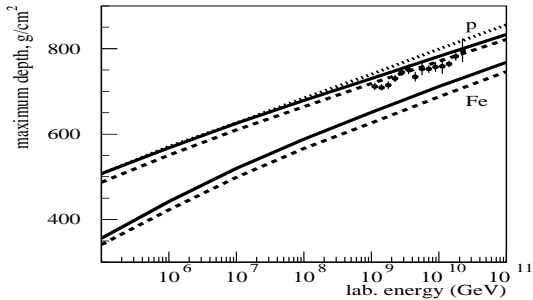} 
\end{center}
\vspace*{-1.3cm}
   \caption{Average position of the shower maximum for proton-
   and iron-induced EAS for QGSJET-II, QGSJET-I,
   and SIBYLL 2.1 models - smooth, dashed, 
   and dotted  curves. The points are the data 
   of the HIRES collaboration \cite{hires}.\label{xmaxair}} 
\end{figure} 

Sizably smaller inelasticities of the new model lead to a somewhat deeper
position of the shower maximum, compared to the original QGSJET -- 
Fig. \ref{xmaxair}. The corresponding shift of $X_{\max}$ changes from
about 20 g/cm$^2$ at $10^{14}$ eV to  $\sim$10 g/cm$^2$ at $10^{20}$ eV
in case of proton-induced EAS; for iron-induced EAS it increases slowly from
15 g/cm$^2$ till $\sim$20 g/cm$^2$. As the relative strength of non-linear 
effects is larger for nucleus-nucleus collisions one observes sizable
deviations from the superposition picture for nucleus-induced air showers.

\begin{figure}[t] 
\begin{center}
  \includegraphics[width=7cm,height=3.9cm,angle=0]{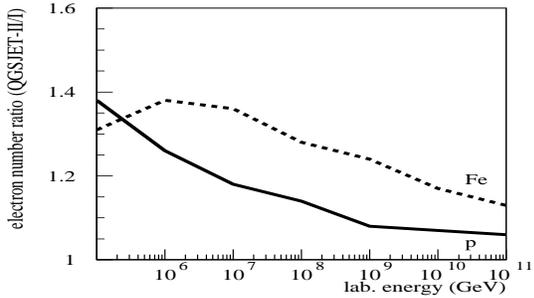} 
\end{center}
\vspace*{-1.3cm}
   \caption{Relative $N_e$-difference at sea level between 
   QGSJET-II and QGSJET-I models for vertical proton-
   and iron-induced EAS -- smooth and dashed curves.
   \label{neair}} 
\end{figure} 

\begin{figure}[t] 
\vspace*{-.4cm}
\begin{center}
  \includegraphics[width=7cm,height=3.9cm,angle=0]{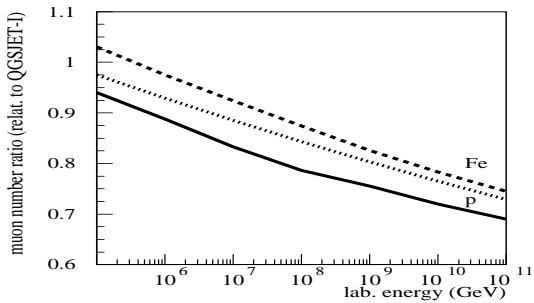} 
\end{center}
\vspace*{-1.3cm}
   \caption{Relative $N_{\mu}$-difference ($E_{\mu}>10$ GeV) at sea level
    between QGSJET-II and QGSJET-I models for vertical proton-
   and iron-induced EAS -- smooth and dashed curves,
   and between SIBYLL 2.1 and QGSJET-I models for  $p-$induced EAS --
   dotted  curve.\label{muair}} 
\end{figure} 

The relative difference between QGSJET-II and QGSJET-I models
 for predicted numbers of electrons and muons 
($E_{\mu}>10$ GeV) at sea level for vertical proton- and iron-induced showers
is shown in Figs.  \ref{neair}, \ref{muair}; 
for comparison the $N_{\mu}$ difference between
SIBYLL 2.1 and  QGSJET-I is also shown for the case of primary protons.
While for electron numbers we observe the expected correlation with the 
position of the shower maximum, the predicted muon numbers in the new model
are significantly smaller, by as much as 30\% at highest energies, due to
a strong reduction of interaction multiplicity, especially for pion-nucleus
interactions, with the results being close to the predictions of the
SIBYLL 2.1 model.

\section{DISCUSSION}

Systematic account for non-linear screening effects in the QGSJET-II model
results in a substantial reduction of multiplicity and inelasticity of
hadron-nucleus interactions compared to the original QGSJET. This leads
in turn to a shift of the predicted position of the shower maximum deeper
in the atmosphere, to  larger number of electrons  and smaller number of
 muons at sea level. Applying the new model to EAS data reconstruction
 will change present conclusions concerning CR composition towards heavier
 primaries. The corresponding $X_{\max}$ change is not large at 
 highest energies. Nevertheless, while the predictions of the original 
 QGSJET, being compared to HIRES measurements, 
 are marginally consistent with the assumption of ultra high energy 
 cosmic rays being only protons, this no longer the case with the new model. 
 
 Similarly, at the knee
 energies QGSJET-II predicts  a substantially smaller number of muons
 for a given number of electrons, which would lead to a conclusion about
 a sizably heavier composition, compared to what is currently obtained with
 QGSJET-I. In particular, the mentioned change should bring closer together
 the composition results of the KASCADE collaboration obtained on the basis
 of electron/muon and hadronic EAS components \cite{hor02}.
 Also the obtained changes seem to go in the right direction to resolve
the reported  discrepancies in the electron/muon correlations \cite{ulr04}. 
It should be mentioned, however, that at highest energies the
 disagreement of the 
composition results obtained with fluorescence light-based measurements 
and with ground arrays \cite{wat04} would increase  using the new model,
as the reduction of EAS muon number at highest energies is much stronger
compared to the corresponding effect for the shower maximum.

It is of interest to investigate the impact of the new model on the
experimental techniques of primary energy reconstruction, where two 
methods are mainly used: either based on the integral of the shower cascade
curve in case of fluorescence light-based measurements or estimating the
energy from charged particle densities at distances 600 and 1000 m from
the shower core \cite{nag00}. In the first case model dependence of the 
results enters mainly via predicted numbers of charged particles in the
shower maximum $N_e^{\max}$, which shows very small model sensitivity --
less than 3\% between QGSJET-II, QGSJET-I, and SIBYLL 2.1. 
Similarly, for electron density at large distances
from the core one observes just a few per cent shift between the new and the
old versions of QGSJET. However, depending on experimental techniques
applied, a substantial contribution to the measured signal may come from the
muon component, which is the case for example, for AUGER collaboration
 \cite{auger}. As the latter is substantially reduced in QGSJET-II,
 corresponding energy estimates of the ground array would move upwards
 compared to QGSJET-I. On the other hand, the possibility to perform
 independent energy reconstructions with both the fluorescence detector
 and the ground array could allow to perform a model consistency check
 at the highest CR energies.

\end{document}